# Direct generation of tunable orbital angular momentum beams in microring lasers with broadband exceptional points

William E. Hayenga,[1†] Jinhan Ren,[1†] Midya Parto,[1†] Fan Wu,[1] Mohammad P. Hokmabadi,[1] Christian Wolff,[2] Ramy El-Ganainy,[3] N. Asger Mortensen,[2,4] Demetrios N. Christodoulides,[1] and Mercedeh Khajavikhan[1*]

[1]CREOL, The College of Optics and Photonics, University of Central Florida, Orlando, Florida 32816-2700, USA

[2] Center for Nano Optics, University of Southern Denmark, Campusvej 55, DK-5230 Odense M, Denmark

[3]Department of Physics and Henes Center for Quantum Phenomena, Michigan Technological University, Houghton, Michigan 49931, USA

[4] Danish Institute for Advanced Study, University of Southern Denmark, Campusvej 55, DK-5230 Odense M, Denmark

[†]These authors contributed equally, [*]Corresponding author: mercedeh@creol.ucf.edu

**Non-Hermitian exceptional points (EPs) represent a special type of degeneracy where not only the eigenvalues coalesce, but also the eigenstates tend to collapse on each other. Recent studies have shown that in the presence of an EP, light-matter interactions are profoundly modified, leading to a host of novel optical phenomena ranging from enhanced sensitivity to chiral light transport. As of now, however, in order to stabilize a system at the vicinity of an exceptional point, its related parameters must be carefully tuned and/or continuously controlled. To overcome this limitation, here we introduce a new family of broadband exceptional points based on unidirectional coupling, implemented by incorporating an S-shaped waveguide in a microring cavity. In active settings, the resulting unidirectionality exhibits unprecedented resilience to perturbations, thus providing a robust and tunable approach for directly generating beams with distinct orbital angular momentum (OAM). This work could open up new possibilities for manipulating OAM degrees of freedom in applications pertaining to telecommunications and quantum computing, while at the same time may expand the notions of non-Hermiticity in the orbital angular momentum space.**

In recent years, there has been a growing interest in the physics and applications of non-Hermitian systems[1,2]. In wave optics, non-Hermiticity, introduced through gain and/or loss, leads to novel mechanisms for light generation and transport. The rapid developments in the field of non-Hermitian physics have so far enabled a host of new functionalities, ranging from unidirectional light propagation and robust chiral mode conversion, to enhanced sensitivity, to name a few[3-10]. Unique to these systems is a special type of degeneracy, known as an exceptional point, at which not only do the eigenvalues coalesce, but also their associated eigenvectors become collinear. To a great extent, the recent advances in the field of non-Hermitian photonics can be attributed to our ability to systematically generate such exceptional points in a number of configurations. In this regard, so far, three archetypical arrangements have been identified and wildly utilized for generating EPs in microcavities. These include parity-time (PT) symmetric photonic molecules[9,11,12], PT-symmetric azimuthal gratings[10], and the incorporation of two Rayleigh scatterers in the periphery of a resonator[8,13].

Angular momentum is an important quantity in physics, its common conservation being intimately connected with rotational symmetry. In optics, beams with orbital angular momentum (OAM), exhibiting intensity nulls at the center and twisted wavefronts, have found numerous applications in microscopy, micromanipulation, quantum computing, fiber optic telecommunications, and sensing[14-20]. Over the years, various strategies have been proposed to convert standard beams like a Gaussian wave into a vortex beam using off-the-shelf and/or custom optical components. For example, one can transform an arbitrary wavefront into a vortex OAM beam by manipulating its transverse amplitude and phase profile using appropriately designed phase plates[21], spatial light modulators, computer generated holograms[22], q-plates[23], or metasurfaces[24]. In integrated settings, a silicon microring resonator with vertical gratings was found to radiate light with lower-order orbital angular momenta when excited through an adjacent bus waveguide[25]. Despite the large body of work on passive approaches for OAM conversion, the quest for directly generating such twisted beams within a laser cavity has just recently begun[26-28].

In this work, we introduce a new type of non-Hermitian photonic arrangement that supports chiral exceptional points. Our proposed structure consists of an active microring resonator with an enclosed S-shaped waveguide with tapered ends. This configuration can be designed such that it operates exclusively in the broken phase (above the exceptional point). We further show that the resulting unidirectionality, that is robust against scatterings off the walls, can be effectively utilized to demonstrate light sources emitting helical beams with a predetermined topological charge. The robustness of this approach is validated by manipulating the OAM order through grating structures and temperature tuning, thus for the first time realizing a tunable orbital angular momentum laser on a III-V semiconductor platform.

By virtue of their geometry and high index contrast, microring resonators support whispering gallery modes (WGMs) that carry orbital angular momentum. The OAM order ($m$) is manifested in the azimuthal phase associated with a traveling wave WGM ($e^{im\phi}$). However, due to the rotation symmetry, WGMs are double degenerate, i.e. an active microring structure simultaneously supports beams of equal positive and negative topological charges, resulting in a net zero OAM generation. Further complications arise when scattering off the cavity walls couples these two counter-propagating modes, resulting in the formation of standing-wave type bi-directional supermodes. Consequently, in order to be able to use microring lasers for vector vortex generation, it is imperative to break the rotation symmetry between the two counter-propagating modes of the ring. Such symmetry breaking will normally manifest in frequency splitting of the two counter-propagating modes, while our exploration of unidirectional coupling offers a new paradigm to unidirectional operation.

## Results

**Establishing a broadband chiral exceptional point via an S-bend in a microring**

To enforce unidirectional operation in active microrings, here we propose to use an S-shaped waveguide with non-reflecting lossy ends inside the resonator. A schematic of this structure with additional corrugations on the exterior cavity walls is depicted in Fig. 1a. The two ends of the S-shaped waveguide are adiabatically tapered in order to ensure energy dissipation (radiation loss) and negligible power reflection. As a result, the S-bend provides asymmetric loss/coupling between the two counter-propagating modes. One can readily show that at resonance, this structure only promotes energy flow in one direction. The asymmetric coupling, caused by the lossy ends of the S-shaped construct, leads to an exceptional point that is inherently chiral[29] with a directionality that can be changed by flipping the cavity design upside-down with respect to the substrate.

One can formally justify the formation of such chiral EPs using temporal coupled-mode theory. In a ring with an S-bend similar to that shown in Fig. 1a, the field amplitudes in the clockwise and counterclockwise directions ($E_{CW}$ & $E_{CCW}$) are related through the following system of coupled equations:

$$\begin{cases} \dot{E}_{CW} = i\omega_0 E_{CW} + (g - \gamma)E_{CW} \\ \dot{E}_{CCW} = i\omega_0 E_{CCW} + (g - \gamma)E_{CCW} + i\mu E_{CW} \end{cases} \quad (1)$$

where $\omega_0$ is the resonant frequency, $g$ is the linear gain, $\gamma$ represents the linear losses due to structural/material losses or cavity decay, and $\mu$ signifies the unidirectional coupling from the clockwise to the counterclockwise component. The detailed derivation of Eq. 1 from first principles as well as the related spatial coupled mode analysis can be found in Supplementary Part 1. Even though in this representation, the associated Hamiltonian attains unequal non-diagonal elements, we emphasize that this arises entirely from combinations of fully reciprocal optical elements and it is by no means an indication of a non-reciprocal behavior. This aspect is discussed further in Supplementary Part 2. To avoid/reduce scattering losses and other unwanted interference effects, the coupling between the S-bend and the ring is provided through proximity. One can then verify that the Hamiltonian associated with this coupled system is non-diagonalizable, featuring an exceptional point with two identical eigenvalues and eigenvectors, given by:

$$\begin{cases} \omega_{1,2} = \omega_0 - i(g - \gamma) \\ \quad |1,2\rangle = (0,1)^T. \end{cases} \quad (2)$$

At the EP, this arrangement robustly supports one of the two counter-propagating modes (here the CCW mode), regardless of the lasing wavelength (this aspect is discussed in Supplementary Part 9). As we will see, this broadband behavior can be advantageously used to facilitate tunable generation of OAM wavefronts of beams. Practically, and to our benefit, any scattering present in the system (intrinsic or intentional) causes some level of coupling between the two counter-propagating modes, which in turn forces the system into the broken phase regime – thus further promoting unidirectionality. This is in sharp contrast with previous reports demonstrating unidirectional behavior[27,28], where a careful adjustment of the parameters is crucial. In other words, in the present configuration, as long as the coupling in one direction is greatly suppressed, the system always remains at the exceptional point. In addition, any additional perturbation that affects both directions in a symmetric fashion has a negligible effect in the resulting unidirectional response. We further validate this robustness by adding a grating on the cavity sidewalls to generate OAM beams with lower orders.

## Generating OAM beams with arbitrary topological charges via sidewall scatterers

In order to down-convert the topological charge carried by the WGM of the microcavity and to promote vertical free-space emission, the following angular phase matching condition $l = p - q$ must be satisfied, where $l\hbar$ is the angular momentum per photon associated with this vector vortex beam, $p$ is the WGM number, and $q$ stands for the number of scattering elements along the periphery (the added grating). An OAM microring laser with a radius of 3 $\mu m$, a width of 500 $nm$ and a height of 210 $nm$, is fabricated on an InP wafer with InGaAsP quantum wells as the gain material. The width of the S-bend is 500 $nm$ and it is located at a nominal distance of 100 $nm$ at its closest proximity to the ring. Square shaped protrusions with sides of 100 $nm$ are incorporated along the outermost sidewall of the ring resonator. These scatterers serve as a second-order grating, generating vertical free-space emission with lower-order twisted beams. Figure 1b displays a scanning electron microscopy (SEM) image of the structure in an intermediate fabrication step. The details of the fabrication can be found in Methods and Supplementary Part 3.

The modal response of such devices is simulated using the finite element methods (FEM). Figure 2a depicts a 2D cross section of the ring waveguide, supporting the transverse electric mode (TE$_0$) which has high overlap with the quantum well structure. Figure 2b shows a top view of the normalized intensity of the electric field in the ring with a tapered S-bend. The uniform intensity of the electric field (as opposed to standard standing wave pattern) is an indication of the unidirectional power flow in the ring. In Fig. 2b, the chirality of the S-bend was selected so as to promote the CCW mode. The mechanism and function of the tapered S-bend in this work is entirely different from previous reports where Y-junction S-bends are used. Unlike a tapered S-bend that leads to an open configuration, rings with Y-junction S-bends feature conservative systems, where interference effects are the primary reason for mode suppression[29]. The OAM conversion process is simulated by including $q = 28$ scattering elements around the periphery of the unidirectional ring (Fig. 2c). Considering the resonance condition at 1540 $nm$ (corresponding to $p = 30$), the vertically extracted field from the scatterers will be given by $\vec{E} \propto \hat{\varphi} e^{il\varphi}$, where $l = 2$. In order to determine the angular momentum of such a vector vortex beam, the corresponding orthogonal left- and right-handed circularly polarized (L,R HCP) components are identified, each having a topological charge of $l - 1 = 1$ and $l + 1 = 3$, respectively, as presented in Fig. 2d. Given these two values, one can then deduce the OAM order of the beam to be equal to $l = 2$.

In order to experimentally measure the topological charge of the microring lasers, a micro-photoluminescence (μ-PL) station with additional branches for measuring the OAM order has been employed. The devices are optically pumped by a fiber laser operating at a wavelength of 1064 $nm$. To accurately measure the OAM beams, the triangular method introduced in[30] has been adopted. The far-field diffraction pattern of the emitted beam is assessed upon passing through a circular polarizer (a combination of a quarter-wave plate and a linear polarizer) as well as an equilateral triangular aperture. This triangle-shaped aperture is placed at the back focal plane of a lens and the InGaAs camera is placed at the front focus. When a beam carrying a non-zero topological charge is incident upon the aperture, a triangular lattice forms that is rotated by $\pm 30°$ compared to the triangle itself. The number of spots along each side is determined by the relationship $s = |l| + 1$. The direction of the rotation denotes the sign of the OAM order. More details on the characterization setup can be found in Methods and Supplementary Part 4. This technique is also compared to the more familiar self-interference approach to establish an equivalence of the two methods. The comparison can be found in Supplementary Part 5.

In the experiments, we first characterized a microring with a tapered S-bend that supports oscillations in the $27^{th}$ ($p = 27$) WGM. In order to be able to measure the OAM order, the microring has been equipped with $q = 26$ scattering elements on its sidewall. These scatterers lower the orbital angular momentum of the beam to $l\hbar$ per photon, where $l = 1$. The intensity profile presented in Fig. 3a displays the doughnut shape associated with a vortex beam. As expected, the intensity profile of the emission is predominantly azimuthally polarized (Fig. 3b). Moreover, using the triangle technique, the OAM characterization indicates a value of $l = 1$, measured by decomposing the beam into two circularly polarized components with topological charges of $l + 1 = 2$ (RHCP) and $l - 1 = 0$ (LHCP), respectively (Fig. 3c-d). Further evidence for unidirectional power flow in rings with tapered S-bends is provided in Supplementary Part 6. A single shot spectrum above threshold is displayed in Fig. 3e, showing a single mode at 1529 $nm$. The light-light curve presented in Fig. 3f, features a threshold behavior characteristic for lasing. Other structures with various radii are fabricated and tested (please see Supplementary Part 7 for some examples).

To further verify the operating principle of the OAM laser, a number of ring resonators are fabricated with the same radii of 3 $\mu m$ ($p = 27$) but with varying number of equally distanced scattering elements ($q = 26, 27, 28$) in their peripheries, resulting in the OAM order $l = +1, 0,$ and $-1$ as depicted in Fig. 4a. The measurement results presented in Fig. 4b clearly confirm the decreasing order of topological charge. Most notably, when the number of grating pitches is equal to the resonant WGM number of the ring, the resulting emission displays a zero net OAM, while remaining an azimuthally polarized vortex beam.

**Tuning the order of the generated OAM beam from a microring laser**

To tune the topological charge of the vector vortex beam, one can use temperature changes. The temperature variations affect the spectral properties of the photoluminescence emission of the gain medium, while at the same time modifying the resonance condition due to a finite thermo-optic coefficient. Here, the order of the emitted WGM is primarily determined from the resulting shift of the photoluminescence radiation. At a fixed pitch number ($q$), the change in the number of WGM order ($p$), directly translates to a change in the topological charge ($l$). In Fig. 5, the temperature tuning of the OAM order has been experimentally demonstrated. At room temperature (295 K) the OAM order of the beam is $l = +1$ (Fig. 5a-b), associated with $p = 27$ and $q = 26$. Once its ambient temperature is reduced to 235 K (blue shifting the photoluminescence peak, hence p= 28 ), the laser emanates a beam with a different OAM, characterized by $l = +2$ (Fig. 5c-d). Yet, at a lower temperature of 150 K, the lasing action occurs at $29^{th}$ WGM, resulting in a topological charge of $l = +3$ (Fig 5e-f). Figure 5g depicts the emission spectra of lasers generating various OAM orders at different temperatures. The characterization results of a larger ring with a radius of 5 $\mu m$, offering an OAM tuning of 4 orders over 75 K temperature range (295K–220K), is reported in Supplementary Part 8.

## Discussion

In conclusion, in this work we demonstrated a new type of exceptional point in micro-resonators, resulting from the incorporation of an enclosed tapered S-shaped waveguide. Due to the asymmetric power loss, the S-bend promotes light propagation in one direction, thus leading to a chiral behavior that is robust against perturbations. Fabricated on an active III-V semiconductor platform, this type of microring structure is used to directly generate vector vortex beams with a pre-determined topological charge. A set of periodically positioned scattering elements along the periphery of the ring imposes an angular phase matching condition and enables efficient vertical free-space emission of vector vortex beams with lower OAM orders. To validate the robustness of this approach, we further tune the topological charge by controlling the temperature. Our work features the first realization of exceptional points with robust and deterministic chirality in microring structures, and may open up new possibilities for sensing, studying non-Hermiticity in angular momentum space, and vortex beam generation.

## Methods

**Fabrication.** The OAM microring laser is fabricated on III-V semiconductor platform. The gain material consists of six InGaAsP quantum wells with an overall height of 200 $nm$ grown on a p-type InP substrate. The quantum wells are covered by a 10 $nm$ thick InP over-layer for protection. Hydrogen silsesquioxane (HSQ) solution in methyl isobutyl ketone (MIBK) is used as a negative tone electron beam resist. The rings are patterned with electron beam lithography, where the exposed HSQ serves as a mask for the subsequent reactive ion etching process. A mixture of $H_2$:$CH_4$:Ar gas chemistry is used with a ratio of 40:10:7 sccm, RIE power of 150 $W$, and ICP power of 150 $W$ at a chamber pressure of 35 $mT$. The wafer is then cleaned with oxygen plasma to remove organic contaminations and polymers that form during the dry etching process. The patterns are submerged in buffered oxide etch (BOE) for 10 seconds to remove the HSQ mask. After this, a 2 $\mu m$ layer of $SiO_2$ is deposited onto the wafer using plasma-enhanced chemical vapor deposition (PECVD). SU-8 3010 photoresist is used to bond the wafer to a glass substrate for mechanical support. Lastly, the remaining InP substrate is completely removed by wet etching in hydrochloric acid.

**Experimental arrangement.** The OAM microring laser is pumped by a pulsed fiber laser operating at a wavelength 1064 $nm$ (15 $ns$ pulse width, 290 $kHz$ repetition rate). The pump beam is focused onto the sample with a 50x objective, this objective in turns also collects the emission from the sample. Light is then either directed to a linear array detector for spectral measurements or to an IR camera for modal profile observation. A triangular aperture is inserted at the back focal plane of the lens before the IR camera for OAM measurements.

## Acknowledgements

National Science Foundation (ECCS 1454531, DMR-1420620, ECCS 1757025), Office of Naval Research (N0001416-1- 2640, N00014-18-1-2347), Air Force Office of Scientific Research (FA9550-14-1-0037), Army Research Office (W911NF-16-1-0013, W911NF-17-1- 0481), U.S.-Israel Binational Science Foundation (BSF) (2016381), DARPA (D18AP00058, HR00111820042, HR00111820038), VILLUM Fonden (16498), and H2020 Marie Skłodowska-Curie Actions (MSCA) (713694).

## Contributions

W.E.H. and M.K. conceived the idea. J.R. fabricated the samples. M.K., J.R., C.W. and N.A.M. developed the related coupled mode theory, and M.P., J.R., F.W. and W.E.H. designed and analyzed the structures. W.E.H. designed the characterization setup and together with J.R. performed the measurements. All authors contributed in interpreting the results and preparing the manuscript.

## Data availability

The datasets generated during and analyzed during this study are available from the corresponding author on reasonable request.

## Additional information

Supplementary Information accompanies this paper at.
Competing interests: The authors declare no competing interests.

# Figures

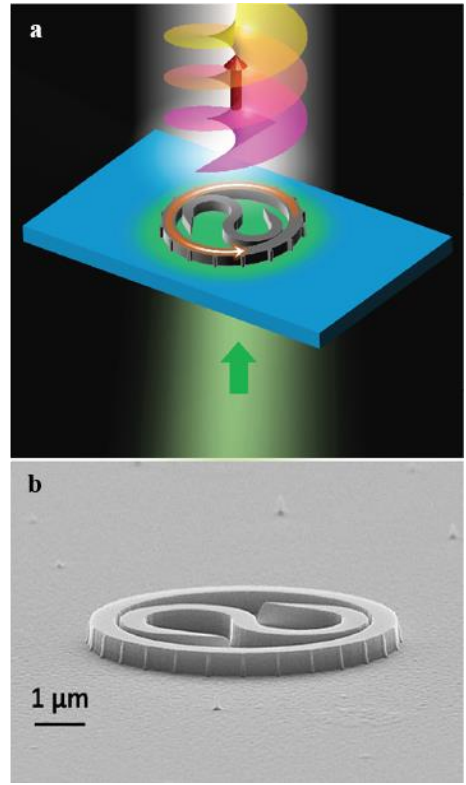

**Fig. 1** OAM microring laser. **a**, Schematic of the OAM microring laser. **b**, SEM image of a fabricated device. The ring has a radius of 3 μm, a width of 0.5 μm and a height of 0.21 μm. Each scatterer has the dimension of 100 nm by 100 nm on the top surface.

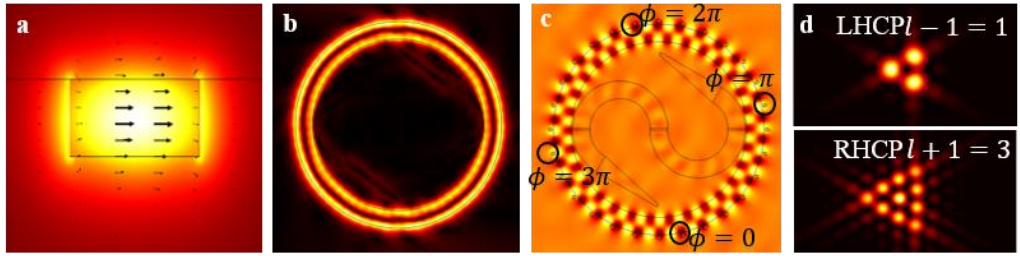

**Fig. 2** Mode analysis of the S-bend ring resonator. **a**, Field intensity of the fundamental TE-mode supported by the cross-sectional waveguide of the microring. **b**, Normalized intensity of the azimuthal component of the E-field inside the structure. **c**, Phase evolution of the scatterers located in the outer periphery of the device when $l = p - q = 2$. **d,** Diffraction patterns of left and right-handed circularly polarized fields through an equilateral triangular aperture.

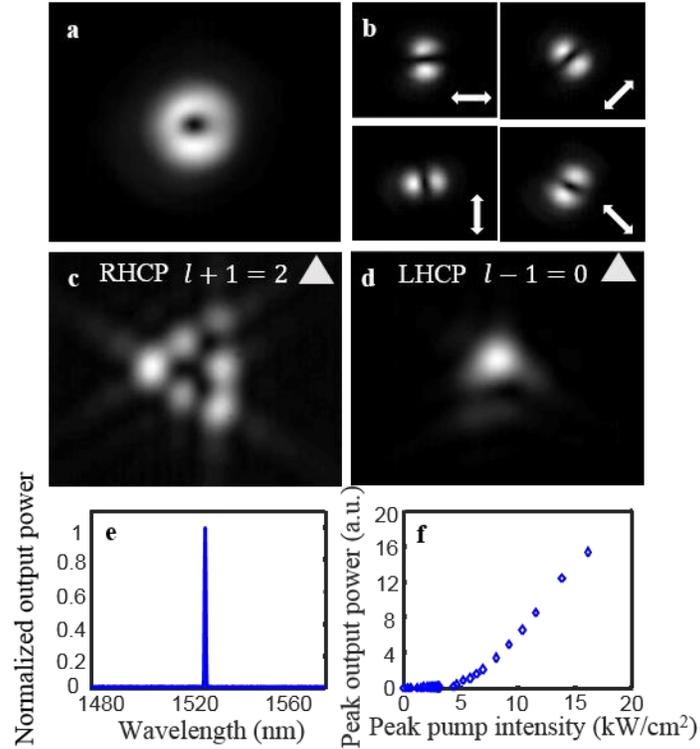

**Fig. 3** Characterization of an OAM laser. **a**, The intensity profile of the microring laser showing a doughnut-shape beam. **b**, Intensity distribution of the lasing emission through a linear polarizer at different orientations (arrows) indicate an azimuthally polarized beam. **c-d**, The far-field diffraction pattern of the emission after passing through an equilateral triangle filtered for the right-hand or left-hand circular polarizations, confirming a vortex beam with $l = +1$. **e**, Single-shot lasing spectrum, confirming single-mode lasing at 1529 nm. **f**, Light-light laser curve, indicating a lasing threshold of 4.1 kW/cm$^2$.

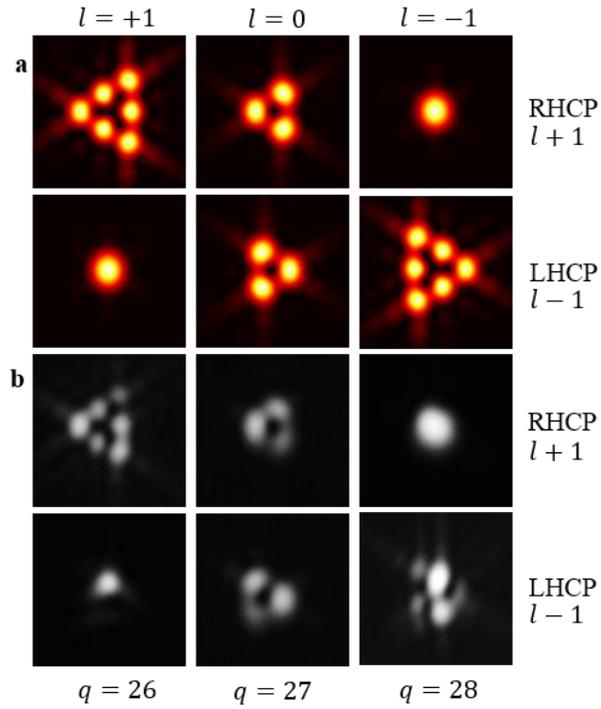

**Fig. 4** OAM lasers with varying number of scattering elements. **a**, Numerical results for the far-field diffraction of a light beam possessing different OAM orders by an equilateral triangle aperture. **b**, The experimental results of the far-field diffraction through the triangle filtered for the RHCP or LHCP, with respect to number of scatterers $q = 26, 27, 28$. In all cases, the lasers are operating in the $p = 27$ WGM.

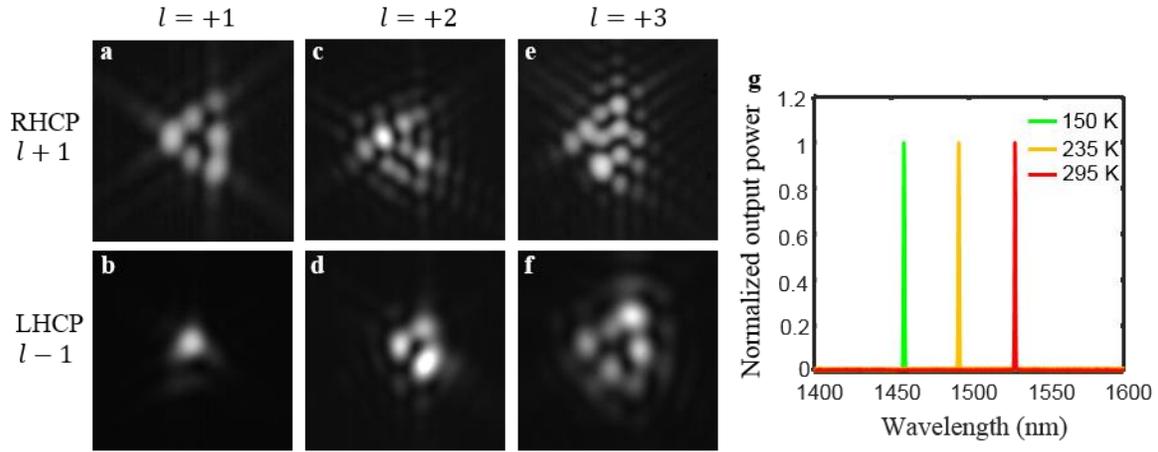

**Fig. 5** Adjusting the OAM of a microring laser with $q = 26$ scattering elements, by temperature tuning the lasing mode. **a-b**, The experimental results for the room temperature laser ($T = 295\ K,\ p = 27$), $l = +1$ is observed. **c-d**, The sample is cooled to $T = 235\ K$, shifting the lasing WGM to the $28^{th}$ mode. Consequently, the OAM order of the light changes to $l = +2$. **e-f**, The laser is further cooled to $T = 150\ K$ ($p = 29$), resulting in an OAM of $l = +3$. **g**, The corresponding spectrum of the laser at varying temperatures.

# Direct generation of tunable orbital angular momentum beams in microring lasers with broadband exceptional points

William E. Hayenga,[1†] Jinhan Ren,[1†] Midya Parto,[1†] Fan Wu,[1] Mohammad P. Hokmabadi,[1] Christian Wolff,[2] Ramy El-Ganainy,[3] N. Asger Mortensen,[2,4] Demetrios N. Christodoulides,[1] and Mercedeh Khajavikhan[1*]

[1]CREOL, The College of Optics and Photonics, University of Central Florida, Orlando, Florida 32816-2700, USA

[2] Center for Nano Optics, University of Southern Denmark, Campusvej 55, DK-5230 Odense M, Denmark

[3]Department of Physics and Henes Center for Quantum Phenomena, Michigan Technological University, Houghton, Michigan 49931, USA

[4] Danish Institute for Advanced Study, University of Southern Denmark, Campusvej 55, DK-5230 Odense M, Denmark

[†]These authors contributed equally, [*]Corresponding author: mercedeh@creol.ucf.edu

## Supplementary Information

### Part 1. Coupled mode analysis of a ring with an S-bend

The Hamiltonian provided in Eq. 1 of the main text can be derived rigorously using either temporal or spatial coupled mode analysis. For the temporal coupled mode analysis, we follow the method used in [1]. In this analysis, each coupling region is considered a port, where $\mathbf{S}^+$ is the input to the resonator and $\mathbf{S}^-$ is the respective output. Figure S1 represents the evolution of the CW and CCW beams. Assuming both coupling sections are the same, one can write the following expressions:

$$\dot{a}_{cw} = (i\omega_0 + g - \gamma)a_{cw} + \kappa S_1^+ + \kappa S_2^+ \qquad \text{Eq. (S1)}$$

$$S_1^- = C S_1^+ + d\, a_{cw} \qquad \text{Eq. (S2)}$$

$$S_2^- = C S_2^+ + d\, a_{cw} \qquad \text{Eq. (S3)}$$

$$\dot{a}_{ccw} = (i\omega_0 + g - \gamma)a_{ccw} + \kappa S_3^+ + \kappa S_4^+ \qquad \text{Eq. (S4)}$$

$$S_3^- = C S_3^+ + d\, a_{ccw} \qquad \text{Eq. (S5)}$$

$$S_4^- = CS_4^+ + da_{ccw} \qquad \text{Eq. (S6)}$$

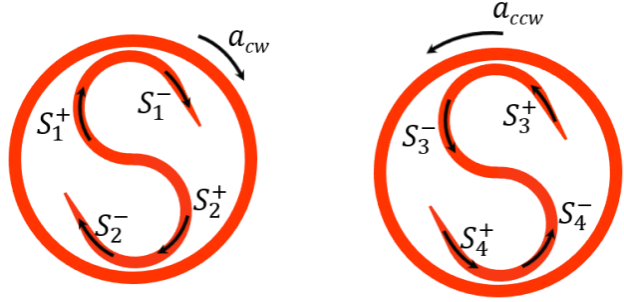

**Fig. S1| Ring resonator with an inner S-bend.** Fields marked as used in the temporal coupled mode analysis.

Here g represents gain and $\gamma$ cavity losses, $S_i^+$ is the input field to port i and $S_i^-$ the field in the corresponding outgoing port. In addition, one can write the following expressions based on the geometry of the problem:

$$S_3^+ = S_4^+ = 0 \qquad \text{Eq. (S7)}$$

$$S_2^+ = S_3^- exp(i\phi) \qquad \text{Eq. (S8)}$$

$$S_1^+ = S_4^- exp(i\phi) \qquad \text{Eq. (S9)}$$

Where $\phi$ is the phase shift the field acquires as it travels along the S-bend from one point of contact to the other. Consequently, from Eqs. (S4) and (S7), one can derive:

$$\dot{a}_{ccw} = (i\omega_0 + g - \gamma)a_{ccw} \qquad \text{Eq. (S10)}$$

While from Eqs. (S5), (S6), and (S7), one arrives at $S_3^- = da_{ccw}$ and $S_4^- = da_{ccw}$. This can be used in conjunction with Eqs. (S1), (S8), and (S9) to get:

$$\dot{a}_{cw} = (i\omega_0 + g - \gamma)a_{cw} + 2\kappa d \exp(i\phi)a_{ccw} \qquad \text{Eq. (S11)}$$

Defining $i\mu = 2\kappa d \exp(i\phi)$, the system of coupled equations in Eq. (S10) and Eq. (S11) is the Hamiltonian equations provided in Eq. (1) of the main text.

One can arrive to a similar system of equations using the standard spatial coupled mode analysis. In this case, one may consider Figure S2.

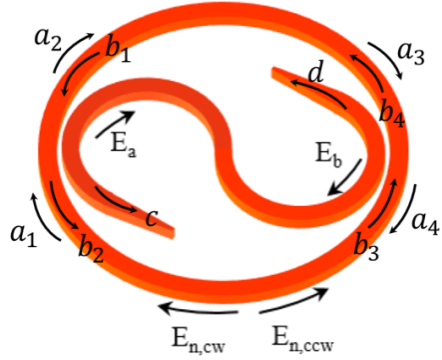

**Fig. S2| Ring resonator with an inner S-bend.** The fields are marked in different sections used in the spatial coupled mode analysis.

The clockwise and counterclockwise field amplitudes of the *n*-th roundtrip are denoted as $E_{n,cw}$ and $E_{n,ccw}$. We assume identical field through- and cross- coupling coefficients, σ and κ, respectively. Then the *n+1*th roundtrip amplitudes is calculated based on the CW and CCW fields in the nth round trip.

$$\begin{bmatrix} a_2 \\ E_a \end{bmatrix} = \begin{bmatrix} \sigma & \kappa \\ \kappa & \sigma \end{bmatrix} \begin{bmatrix} a_1 \\ 0 \end{bmatrix} \quad \text{Eq. (S12)}$$

$$\begin{bmatrix} b_2 \\ c \end{bmatrix} = \begin{bmatrix} \sigma & \kappa \\ \kappa & \sigma \end{bmatrix} \begin{bmatrix} b_1 \\ E_b g_s e^{-i\phi_s} \end{bmatrix} \quad \text{Eq. (S13)}$$

$$\begin{bmatrix} a_4 \\ E_b \end{bmatrix} = \begin{bmatrix} \sigma & \kappa \\ \kappa & \sigma \end{bmatrix} \begin{bmatrix} a_3 \\ 0 \end{bmatrix} \quad \text{Eq. (S14)}$$

$$\begin{bmatrix} b_4 \\ d \end{bmatrix} = \begin{bmatrix} \sigma & \kappa \\ \kappa & \sigma \end{bmatrix} \begin{bmatrix} b_3 \\ E_a g_s e^{-i\phi_s} \end{bmatrix} \quad \text{Eq. (S15)}$$

Where σ and κ are the through- and cross- coupling coefficients, respectively. The electrical field of $c$, and $d$ will vanish eventually due to the radiation loss of the tapered S-bend. One can easily find the relation of the electrical field between $a_2$ and $a_1$, $b_2$ and $b_1$, $a_4$ and $a_3$, $b_4$ and $b_3$:

$$a_2 = \sigma a_1 \quad \text{Eq. (S16)}$$

$$b_2 = \sigma b_1 + \kappa E_b g_s e^{-i\phi_s} \qquad \text{Eq. (S17)}$$

$$a_4 = \sigma a_3 \qquad \text{Eq. (S18)}$$

$$b_4 = \sigma b_3 + \kappa E_a g_s e^{-i\phi_s} \qquad \text{Eq. (S19)}$$

And $a_3$ and $a_2$, $b_1$ and $b_4$ have the relation of:

$$a_3 = g_r^{\frac{1}{2}} e^{-\frac{i\phi_r}{2}} a_2 \qquad \text{Eq. (S20)}$$

$$b_1 = g_r^{\frac{1}{2}} e^{-\frac{i\phi_r}{2}} b_4 \qquad \text{Eq. (S21)}$$

also,

$$E_{n+1,cw} = g_r^{\frac{1}{4}} e^{-\frac{i\phi_r}{4}} a_4 \qquad \text{Eq. (S22)}$$

$$a_1 = g_r^{\frac{1}{4}} e^{-\frac{i\phi_r}{4}} E_{n,cw} \qquad \text{Eq. (S23)}$$

$$E_{n+1,ccw} = g_r^{\frac{1}{4}} e^{-\frac{i\phi_r}{4}} b_2 \qquad \text{Eq. (S24)}$$

$$b_3 = g_r^{\frac{1}{4}} e^{-\frac{i\phi_r}{4}} E_{n,ccw} \qquad \text{Eq. (S25)}$$

Substituting Eqs. (S16), (S18), (S20), (S23) into Eq. (S22), one can directly find,

$$E_{n+1,cw} = g_r^{\frac{1}{4}} e^{-\frac{i\phi_r}{4}} a_4 = \sigma^2 g_r e^{-i\phi_r} E_{n,cw} \qquad \text{Eq. (S26)}$$

Similarly, one can also substitute Eqs. (S17), (S19), (S21) and (S25) into Eq. (S24) in order to find the expression of $E_{n+1,ccw}$, which has a form of:

$$E_{n+1,ccw} = \sigma^2 g_r e^{-i\phi_r} E_{n,ccw} + \kappa g_s e^{-i\phi_s} (\sigma g_r^{\frac{3}{4}} e^{-\frac{i3\phi_r}{4}} E_a + g_r^{\frac{1}{4}} e^{-\frac{i\phi_r}{4}} E_b) \qquad \text{Eq. (S27)}$$

Where $E_a = \kappa a_1 = \kappa g_r^{\frac{1}{4}} e^{-\frac{i\phi_r}{4}} E_{n,cw}$, and $E_b = \kappa a_3 = \kappa \sigma g_r^{\frac{3}{4}} e^{-\frac{i3\phi_r}{4}} E_{n,cw}$.

Combining Eq. (S26) and (S27), one can find the coupling matrix for this system as

$$\begin{bmatrix} E_{n+1,cw} \\ E_{n+1,ccw} \end{bmatrix} = \begin{bmatrix} \sigma^2 g_r e^{-i\phi_r} & 0 \\ 2\kappa^2 \sigma g_s g_r e^{-i\phi_s} e^{-i\phi_r} & \sigma^2 g_r e^{-i\phi_r} \end{bmatrix} \begin{bmatrix} E_{n,cw} \\ E_{n,ccw} \end{bmatrix} \qquad \text{Eq. (S28)}$$

**Part 2. Reciprocity and energy conservation in a passive system**

The main text considers a system described by an equation of the form

$$i\partial_t \psi(t) = H\psi(t)$$

$$= i\partial_t \begin{pmatrix} E_{CW}(t) \\ E_{CCW}(t) \end{pmatrix} = \begin{pmatrix} \tilde{\omega} & \kappa_1 \\ \kappa_2 & \tilde{\omega} \end{pmatrix} \begin{pmatrix} E_{CW}(t) \\ E_{CCW}(t) \end{pmatrix}, \qquad \text{Eq. (S29)}$$

of a perturbed ring resonator for the special case of $\kappa_1 = 0$. Here, $E_{CW}(t)$ and $E_{CCW}(t)$ are the clockwise and counterclockwise propagating amplitudes and $\tilde{\omega} = \omega + i\gamma$ takes the form of a complex angular frequency. Naively, one would expect the coupling matrix to be Hermitian or even symmetric. In fact, this is precisely what one obtains when starting with a perfect ring resonator, adding a dielectric perturbation (e.g. the S-bend waveguide studied in the main text) and computing the mode overlap integrals within perturbation theory. One will find $\kappa_1 = \kappa_2^*$ and might incorrectly relate this symmetry to reciprocity, because of the similarity to the symmetry of scattering matrices. In contrast, the Hermicity of $H$ (being the Hamiltonian in a Schrödinger-type equation) is related to the conservation of energy; a fact easily missed in standard coupled-mode perturbation theory.

Nonetheless, this picture is useful to gain deeper understanding: the off-diagonal elements of $H$ describe the scattering of one optical mode into the counter-propagating one, i.e. the *reflection* by the perturbation. This is not constrained at all by reciprocity but only by the conservation of energy. As a result, systems with $\kappa_1 \neq \kappa_2$ are entirely possible as long as they are sufficiently lossy, typically due to scattering into the far-field as a side-effect of the dielectric perturbation. There is not even a fundamental argument against the extreme case of $\kappa_1 = 0$, which leads to dynamics that correspond to an exceptional point as discussed in the main text.

Finally, we now address the question what "sufficiently lossy" means in practice. Although the overall conservation of energy does not require any symmetry of the matrix, it still imposes limits on the values of the parameters $\kappa_2$ and $\widetilde{\omega}$ assuming a *passive* system. Note that this does not imply that no optical pumping is present, but only that it does not overcompensate the intrinsic material loss at any point in space. In a passive system, the overall energy

$$U_{tot} = U_0(|E_{CW}|^2 + |E_{CCW}|^2), \qquad \text{Eq. (S30)}$$

cannot grow in time (here, $U_0$ is the energy of the optical mode). The Schrödinger-type equation (S29) with $\kappa_1 = 0$ can be solved using a Neuman series:

$$\psi(t) = e^{-i\widetilde{\omega}t}\begin{pmatrix} 1 & 0 \\ -i\kappa_2 t & 1 \end{pmatrix}\psi(0) \qquad \text{Eq. (S31)}$$

For the initial state $\psi(0) = [1, 0]^T$, this leads to the time-evolution of the total energy:

$$U_{tot}(t) = U_0|\psi(t)|^2 = U_0(1 + \kappa_2^2 t^2)e^{-2\gamma t} \qquad \text{Eq. (S32)}$$

For times $\left(1 - \sqrt{1 - \left(\frac{2\gamma}{|\kappa|}\right)^2}\right)/2\gamma \leq t \leq \left(1 + \sqrt{1 - \left(\frac{2\gamma}{|\kappa|}\right)^2}\right)/2\gamma$, this function grows, i.e. describes a gain of energy that is unphysical for a passive system. Therefore, we can conclude that a passive ring-like system that is governed by a coupled-mode equation of the form

$$i\partial_t \psi(t) = \begin{pmatrix} \omega + i\gamma & 0 \\ \kappa & \omega + i\gamma \end{pmatrix}\psi(t) \qquad \text{Eq. (S33)}$$

must satisfy the condition $2\gamma > |\kappa|$, which is a lower bound for the optical loss. Of course, this condition no longer applies if net gain is achieved anywhere in the system, e.g. in the laser studied in the main text.

**Part 3. Fabrication**

The fabrication procedure for optically pumped dented microring resonators with an incorporated S-bend is illustrated in Fig. S3. The gain material consists of six quantum wells of $In_{x=0.734}Ga_{1-x}As_{y=0.57}P_{1-y}$ (thickness: 10 nm), each sandwiched between two cladding layers of $In_{x=0.56}Ga_{1-x}As_{y=0.938}P_{1-y}$ (thickness: 20 nm), with an overall height of 200 nm grown on p-type InP substrate. The quantum wells are covered by a 10 nm thick InP over-layer for protection. An XR-1541 hydrogen silsesquioxane (HSQ) solution in methyl isobutyl ketone (MIBK) is used as a negative electron beam resist. The resist is spun onto the wafer, resulting in a thickness of 100 nm, after which it is soft baked (Fig. S3a). The rings are then patterned by electron beam lithography (Fig. S3b). The wafer is next immersed in tetramethylammonium hydroxide (TMAH) for 30 seconds to develop the patterns and rinsed in isopropyl alcohol (IPA) for 30 seconds. The HSQ exposed to the electron beam now remains and serves as a mask for the subsequent reactive ion etching process. To perform the dry etching, a mixture of $H_2$:$CH_4$:Ar gas chemistry is used with a ratio of 40:10:7 sccm, RIE power of 150 W, and ICP power of 150 W at a chamber pressure of 35 mT (Fig. S3c). The wafer is then cleaned with oxygen plasma to remove organic contaminations and polymers that form during the dry etching process. A 50 sccm flow of $O_2$ is used with an RIE power of 150 W and ICP power of 150 W at a chamber pressure of 50 mT. The patterns are then submerged in buffered oxide etch (BOE) for 10 seconds to remove the HSQ mask (Fig. S3d). After this, a 2 μm layer of $SiO_2$ is deposited onto the wafer using plasma-enhanced chemical vapor deposition (PECVD) (Fig. S3e). SU-8 3010 photoresist is used to bond the wafer to a glass substrate for mechanical support (Fig. S3f). After spinning on the photoresist, the wafer is placed onto the glass, pattern-side-down, and exposed for 15 seconds on both sides. Lastly, the remaining InP substrate is completely removed by wet etching in hydrochloric acid (HCl) for 60 min [2] (Fig. S3g).

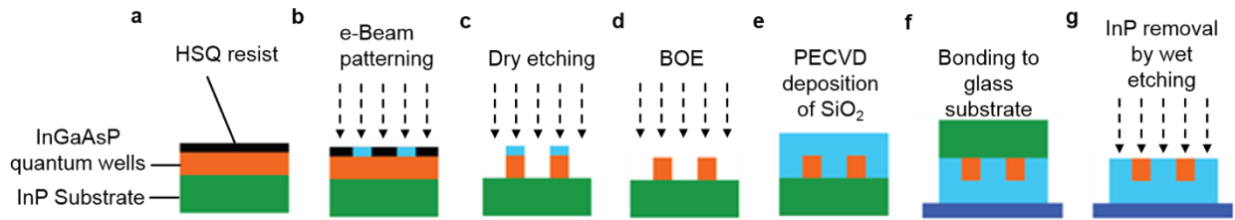

**Fig. S3| Schematic of the fabrication procedure of microring lasers. a**, HSQ e-beam resist is spun onto the wafer. **b**, The wafer is patterned by e-beam lithography. **c**, A dry etching process to define the rings. **d**, The sample is immersed in BOE to remove the masking HSQ. **e**, A 2 $\mu m$ layer of $SiO_2$ is deposited via PECVD. **f**, The wafer is flipped upside-down and bonded to a glass substrate by SU-8 photoresist to provide mechanical support. **g**, Lastly, the InP substrate is wet etched by HCl.

**Part 4. Characterization setup**

A micro-photoluminescence (µ-PL) setup, depicted in Fig. S4, is used to characterize the microring resonators. The microrings are optically pumped by a pulsed (duration: 15 ns, repetition rate: 290 kHz) laser operating at a wavelength of 1064 nm (SPI fiber laser). A beam shaping system is implemented to realize the desired pump size. In this study, the pump has a diameter of ~40 $\mu m$. A $50x$ microscope objective (NA: 0.42) is used to project the pump beam on the ring resonator and also serves to collect the emission. For temperature tuning, the sample is inserted into a cryostat (Janis ST-500) and cooled. The surface of the sample is imaged by two cascaded 4-f imaging systems in an IR camera (Xenics Inc.). A broadband ASE source passed through rotating ground glass is used to illuminate the sample surface for pattern identification. A notch filter is placed in the path of emission to attenuate the pump beam. Output spectra are obtained by a monochromator equipped with an attached linear array InGaAs detector. A powermeter is inserted at the focus of the beam to collect the output power from the microrings. A linear polarizer is placed in the setup to observe the polarization resolved intensity distribution. To measure the OAM, a removable equilateral triangular aperture is inserted at the back focal plan of the lens before the IR camera to facilitate a Fourier-transform. A quarter-wave plate and linear polarizer filter light for different handedness of the polarizations. Additionally, a quasi-parallel Mach-Zehnder interferometer has been built to characterize the OAM of the emission. A more detailed explanation of the triangular aperture method and a comparison to the self-interference technique is provided in Supplementary Part 5.

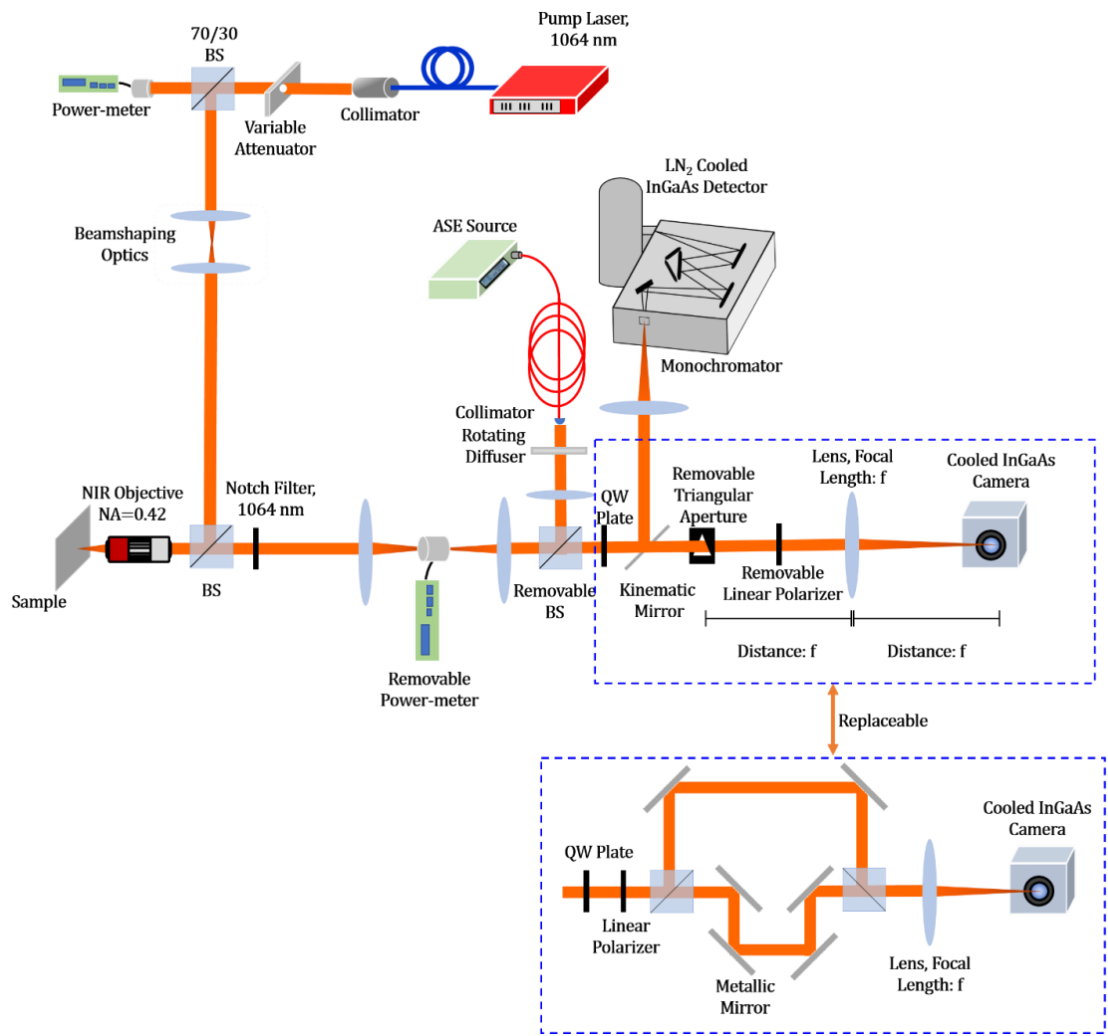

**Fig. S4| Schematic of the μ-PL characterization setup.** The microrings are pumped by a pulsed laser (15 $ns$ pulse width, 290 $kHz$ repetition rate). The pump beam is focused onto the sample with a 50x objective, this objective in turns also collects the emission from the samples. Light is then either directed to a linear array detector for spectral measurements or to an IR camera for modal profile observation. A triangular aperture is inserted at the back focal plane of the lens before the IR camera for OAM measurements. A Mach-Zehnder interferometer operating in quasi-parallel fashion is also incorporated into the setup in order to show the equivalence of the two OAM assessment approaches.

**Part 5. Comparison of two orbital angular momentum detecting techniques**

To measure the orbital angular momentum of light, different techniques have been developed based on interference [3, 4] including self-interference and interference of the OAM beam with a plane wave. In these circumstances, the phase information is assessed by analyzing the fringes. In particular, the formation of a fork-like structure at the center of the vortex is observed. More recently, another approach has been demonstrated that is based on the relationship between the phase of the light carrying OAM and diffraction phenomena [5]. This technique provides an unambiguous measurement of the order and sign of a vortex beam's topological charge. In short, when a *scalar* light beam carrying orbital angular momentum passes through an equilateral triangular aperture, the beam diffracts, hence generating a truncated triangular optical lattice rotated by $\pm 30°$, with respect to the aperture, in the far-field. This lattice then reveals the value of the topological charge ($l$), given by the relationship $|l| = s-1$, where $s$ is the number of spots along each side of the formed triangular lattice. The sign of the charge is determined by the direction that the triangle rotates. For example, in our setup, a triangle pointed left has a positive charge, while the sign is negative if it points right.

To establish the validity of the triangle approach for measuring OAM of our laser, we compare the results of both the triangle technique and the self-interference method, using a unidirectional active microring vortex laser as the source. The dimensions of the ring are the same as described in Fig. 1 of the main text (radius: 3 $\mu m$). The structure has 26 periodic scattering elements along the periphery. Figure S5 presents the comparison results. The emission of the beam displays the doughnut shape associated with a vortex beam (Fig. S5a). The lasing spectrum is provided in Fig. S5b, showing a single mode at $1529\ nm$, corresponding to the $27^{th}$ WGM of the ring resonator. In this study, as discussed in the main text, the light emitted by the unidirectional microring laser is of the form $\vec{E} \propto \hat{\varphi}e^{il\varphi} = \hat{L}e^{i(l-1)\varphi} + \hat{R}e^{i(l+1)\varphi}$, where $\hat{L}$ and $\hat{R}$ represent left- and right-handed circular polarization components (LHCP, RHCP), respectively. Thus, inserting a quarter-wave plate converts the left- and right-handed circular polarizations to linear horizontal and vertical polarizations. Therefore, a linear polarizer allows for the filtering of one of the topological charges, depending on if it is oriented in the horizontal or vertical direction. From the above analysis, it is expected that for light carrying $l = 1$, topological charges of $l + 1 = 2$ and $l - 1 = 0$ are to be observed when selecting the RHCP or LHCP, respectively. The OAM measurements in Fig. S5c-d (triangle method) and Fig. S5e-f (self-interference) indicate corresponding measurements of topological charges $l + 1 = 2$ (RHCP) and $l - 1 = 0$ (LHCP). From these results, it follows that the original vector vortex beam has an angular momentum of $l\hbar$ per photon, with $l = 1$. This confirms the equivalence of the two OAM characterization schemes.

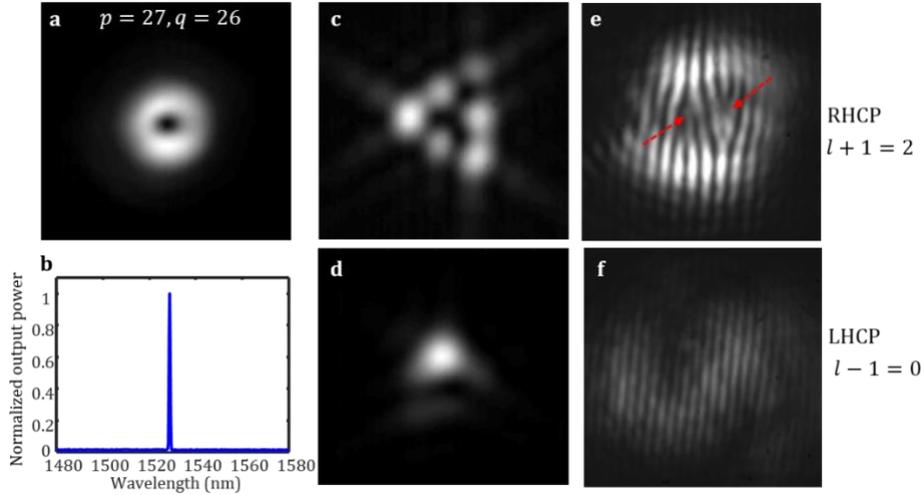

**Fig. S5| OAM characterization equivalence between the triangular aperture and self-interference methods applied to a unidirectional active microring vortex laser. a**, The doughnut shaped intensity profile of the microring laser (radius 3 $\mu m$). **b**, Corresponding spectrum, showing a single mode at 1529 $nm$. **c-d** The far-field diffraction patterns of the vector vortex beam ($l = 1$) after passing through the triangular aperture and filtering for the scalar RHCP ($l + 1 = 2$) or LHCP ($l - 1 = 0$) components. **e-f**, Self-interference of the vortex beam of the two filtered circular polarizations. The RHCP displays two 3-pronged forks pointed in opposite directions (marked by arrows) located at the centers of the two beams, indicating a topological charge of $2 = l + 1$. The LHCP shows uniform fringes, confirming the observations assessed by the triangle far-field diffraction pattern.

**Part 6. Unidirectional propagation in a microring resonator**

Microring resonators, in principle, tend to lase in both clockwise (CW) and counterclockwise (CCW) modes. In most cases, however, the roughness of the waveguide walls couples these two counter-propagating modes, leading to the formation of a set of bi-directional supermodes. To enforce the unidirectional power flow, a ring resonator with an S-shaped chiral element is utilized as illustrated in Fig. S6a (as opposed to a standard ring without an S-bend in Fig. S6b). The S-bend breaks the chiral symmetry of the ring by allowing coupling from the CW to the CCW propagation, while prohibiting the energy exchange in the opposite direction. In the presence of gain, the mode experiencing higher coupling loss will be effectively suppressed (in this case the CW mode). To experimentally verify this unidirectional operation, we fabricated ring resonators (radius: **20 μm** width: **500 nm**, height: **210 nm**) with/without an embedded S-bend and a bus-waveguide terminated by second-order gratings. Figures S6c-f compare the light emission in rings with and without S-bend. Both rings are optically pumped at the same power level. Clearly, the light generation from the microring with the S-bend is unidirectional. Measuring the total emitted power from each grating shows a more than $28\ dB$ extinction ratio between the two counter propagating modes (in the favor of CCW mode) in the microcavity with the S-bend (Fig. S6e). In comparison, a ring without the S-bend exhibits less than $0.1\ dB$ power difference between the two terminals (Fig. S6f) [6]. The two device configurations (with and without the S-bend) show similar threshold intensity behaviors.

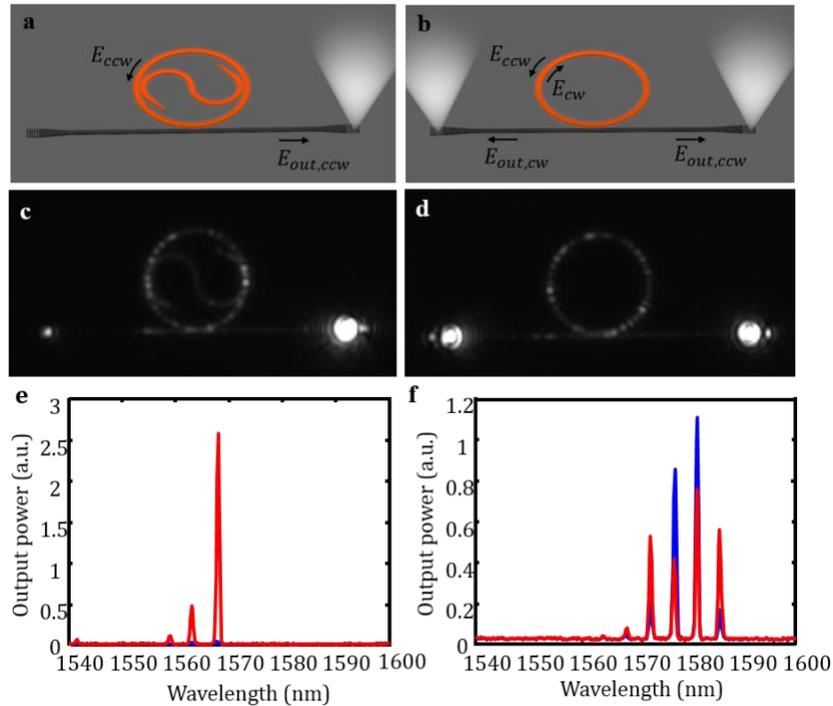

**Fig. S6| Unidirectional emission in microring lasers. a-b**, Schematic of a ring resonator with/without an S-bend. **c-f**, Recorded camera images and spectra from the out-coupling gratings for **c, e,** cavities with the S-bend and **d, f** rings without the S-bend, exhibiting a **28 dB** extinction ratio between the two counter-propagating modes (in this case in favor of the CCW mode).

**Part 7. Measurements of other rings**

Additional ring lasers with the sidewall gratings and the S-bends having radii of 5 μm and 10 μm (width: 500 nm), are fabricated and characterized. The light-light curves of these structures displayed in Fig. S7 show a reduction of threshold intensity as the radius of the ring increases.

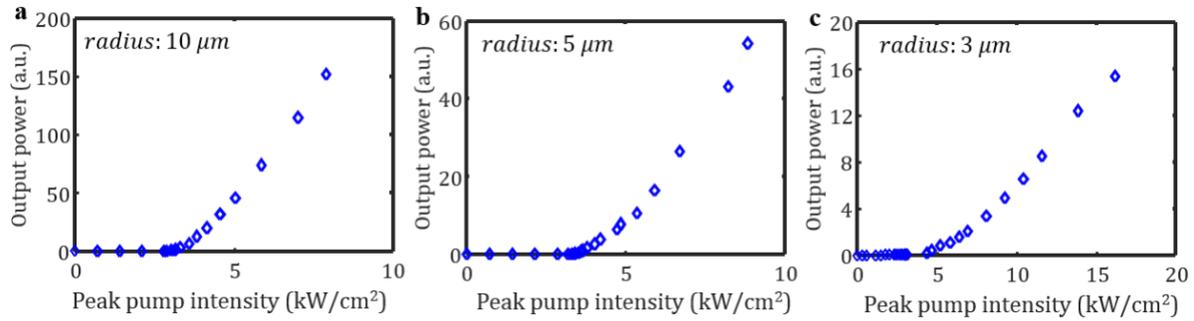

**Fig. S7|  Light-light curve of ring lasers with different radii**: **a**, **10 μm**, **b**, **5 μm** and **c**, **3 μm**. As the size of the ring is reduced the threshold density increases due to bending losses.

**Part 8. Orbital angular momentum temperature tuning**

In larger cavities, due to the increased cavity length, the free spectral range (FSR) is correspondingly smaller. Consequently, changing the order of the mode requires less temperature variations. Here, we present the OAM tuning in a cavity with a radius of 5 $\mu m$ (width: 500 $nm$, gratings: $q = 48$). Figure S8 provides the far-field diffraction pattern through a triangular aperture and the single shot spectra of this structure. At room temperature the ring laser has a topological charge of $l = -5$ (Fig. S8a). As the temperature decreases, the whispering gallery mode hops, increasing the OAM order to $l = -4$ at 280 K (Fig. S8b), $l = -3$ at 245 K (Fig. S8c), and $l = -2$ at 220 K (Fig. S8d). The single shot spectra at each temperature is shown in Fig. S8e.

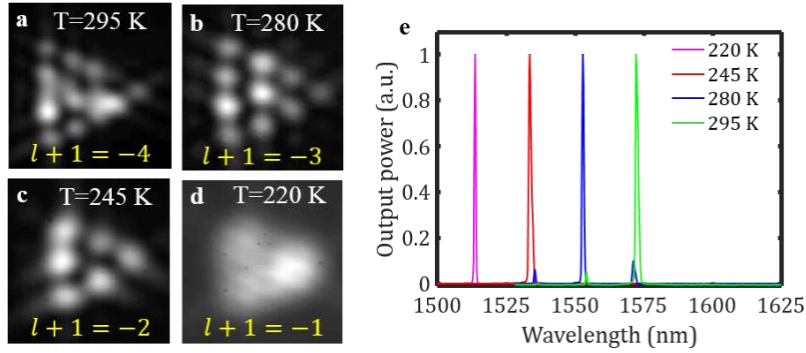

**Fig. S8| The OAM of a temperature tuned 5 $\mu m$ radius microring is observed.** The topological charge measurements associated with the RHCP at different temperatures are shown in **a**, $l = -5$ at 295 K, **b**, $l = -4$ at 280 K, **c**, $l = -3$ at 245 K and **d**, $l = -2$ at 220 K. **e**, Single shot spectra measured at each temperature.

Microrings support whispering gallery modes (WGM) that inherently have large azimuthally varying phases of the form $\vec{E} \propto \hat{\varphi}\, e^{\pm ip\varphi}$, where $p$ is the azimuthal number of a particular WGM, and $\varphi$ is the angular coordinate. Moreover, an incorporation of the S-bend promotes unidirecitonality, and the addition of the gratings allow for a down conversion of the intrinsic OAM of the ring ($p$) to a smaller value. Ultimately, a phase matching condition results in a vertically outcoupled OAM emission of $l = p - q$, where $l$ is the down-converted OAM and $q$ is the number of scattering elements around the ring. One can readily observe that a change in $p$ or $q$ will result in a change in the OAM state.

In order to temperature tune the OAM of the microring, the lasing WGM is adjusted to a new azimuthal mode $p$. To elucidate further on this technique, we first measure the spectral location of the peak of the photoluminescence at various temperatures of the bare InGaAsP wafer used for our structure. Shown below, we observed a change of approximately $0.5 nm/K$ in the maximum of the gain when altering the temperature.

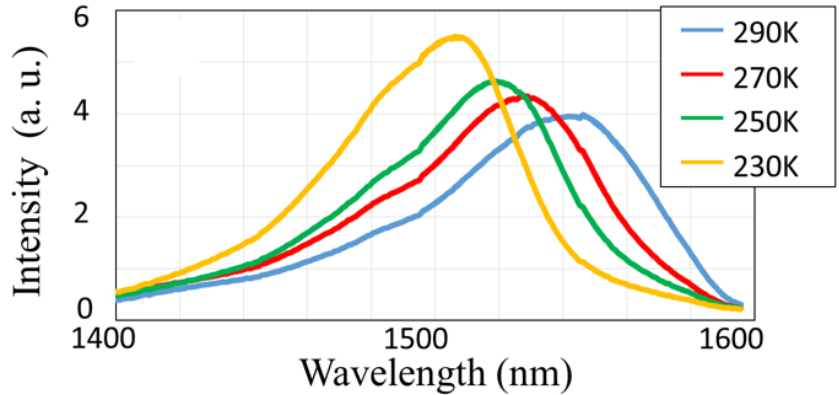

**Fig. S9| Thermal dependence of the QW gain.** A change of approximately 0.5nm/K in the peak of the gain when altering the temperature.

Additionally, shown in Fig. S10 (**a**) is the single-shot spectra of the 3 $\mu m$ radius ring under study in the main text ($q = 26$), taken at three separate temperatures. A ring of this size is expected to exhibit a free spectral range of ~35 $nm$. We observe that at a temperature of 295 $K$, the ring is lasing in the WGM with $p = 27$ (**a,b,c**) and outcouples OAM with $l = 1$. When reducing the temperature by 60 $K$, the change in the photoluminescence peak causes a shift in the resonance wavelength by 35 $nm$, indicating that the structure has shifted from the $p = 27$ to the $p = 28$ WGM (**a**). In turn, this is confirmed by measuring the OAM which is tuned to $l = 2$ (**d,e**) from the initial state of $l = 1$ Further reductions in temperature adjusts the gain peak and the resonant WGM, as shown at 150 $K$, where $p$ is now 29 and $l$ is 3 (**a,f,g**).

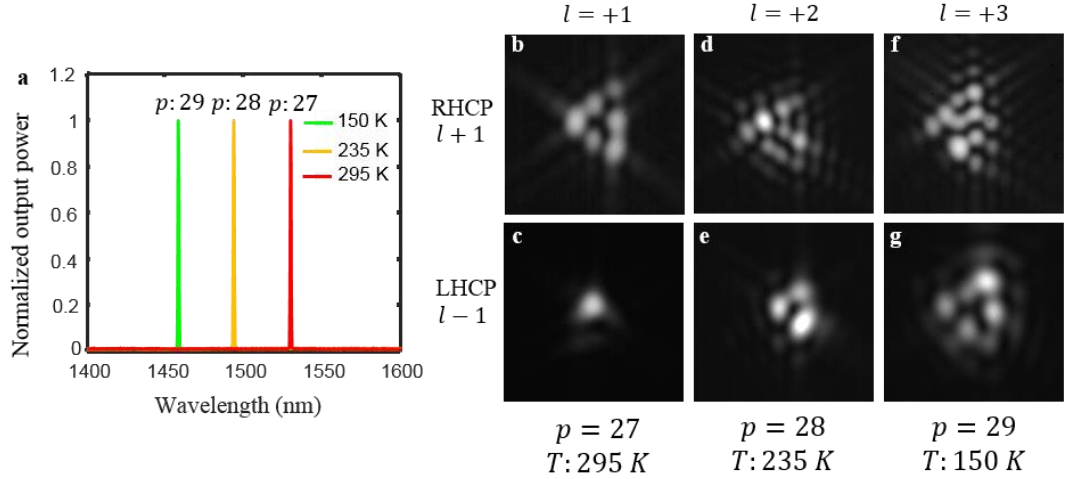

**Fig. S10| Tunability of the OAM of the microring laser.** **a**, The corresponding spectrum of the laser at varying temperatures, detailing the whispering gallery mode number $p$. **b-g**, The experimental results for the room temperature laser ($T = 295\ K$, $p = 27$), $l = +1$ is observed. The sample is cooled to $T = 235\ K$, shifting the lasing WGM to the $28^{th}$ mode. Consequently, the OAM order of the light changes to $l = +2$. The laser is further cooled to $T = 150\ K$ ($p = 29$), resulting in an OAM of $l = +3$.

## Part 9. Broadband unidirectionality and robustness analysis

The unidirectional behavior resulting due the presence of an S-bend is robust against scattering off the walls of the waveguide and can be designed to be broadband across several cavity modes. This allows for the generation of high-fidelity tunable OAM beams. In this section, we provide an analysis of the modes of the microring with a tapered S-bend, in the presence of scattering off the cavity walls that tend to couple the CW and CCW modes.

As derived from the first principles, through both temporal and spatial coupled mode analysis (provided in part 1), the Hamiltonian representing a ring with an S-bend is as follows:

$$\begin{bmatrix} \dot{a} \\ \dot{b} \end{bmatrix} = \begin{bmatrix} 0 & 0 \\ \kappa' e^{i\phi} & 0 \end{bmatrix} \begin{bmatrix} a \\ b \end{bmatrix}, \qquad \text{Eq. (S34)}$$

where $a$ and $b$ are the field amplitudes in the clockwise (CW) and counter clockwise (CCW) directions respectively, $\kappa'$ is coupling between the ring and the S-bend waveguide, and $\phi$ is the optical phase imparted by the S-bend. For mathematical convenience and without loss of generality

(i.e. through a trivial gauge transformation), here we remove the terms presenting the onsite resonance frequencies from the diagonal elements. Clearly, the above matrix is non-diagonalizable. In fact, it is easy to show that the above Hamiltonian is presenting a system exactly at an exceptional point. In this case, the eigenvalue is $\lambda = 0$, and the eigenvector is $[0 \ 1]^T$. It is also easy to verify that the amount of phase accumulation $\phi$ (representing the length of the S-bend) has no effect on the modes of the system. As a result, the exceptional point can occur at any wavelength – as long as the other non-diagonal term is exactly zero.

Of course, assuming perfectly zero coupling from one side to the other is not realistic. In addition, the OAM down-converting grating will add additional scattering off the walls. Here we study the effect of this inevitable/additional scattering on the cavity modes by adding a perturbation of $\epsilon$ to the non-diagonal terms.

$$\begin{bmatrix} \dot{a} \\ \dot{b} \end{bmatrix} = \begin{bmatrix} 0 & \epsilon \\ \kappa' e^{i\phi} + \epsilon & 0 \end{bmatrix} \begin{bmatrix} a \\ b \end{bmatrix}. \qquad \text{Eq. (S35)}$$

The eigenvalues and eigenvectors of the perturbed Hamiltonian are as follows:

$$\lambda_{1,2} = \pm \kappa' \sqrt{(\epsilon/\kappa')^2 + (\epsilon/\kappa')e^{i\phi}}$$
$$V_{1,2} = \left[ 1, \pm \sqrt{\epsilon^2 + \epsilon \kappa' e^{i\phi}}/\epsilon \right]^T \qquad \text{Eq. (S36)}$$

Figure S11 (**a**) shows the imaginary part of the eigenvalues $\lambda_{1,2}$ for different values of the normalized perturbation ($\epsilon/\kappa'$), when $\phi = \pi/2$. As it is evident from this diagram, regardless of the sign of the perturbation $\epsilon$, one of the two respective eigenmodes $V_{1,2}$ will be suppressed due to a lower Q-factor. In addition, whether $V_1$ or $V_2$ prevail, according to Eq. (S36), always a chiral component (in this case CCW) will predominantly lase in the microring structure. This is also evident from the power ratio between these components in the eigenvectors $V_{1,2}$, as shown in Fig. S11(**b**).

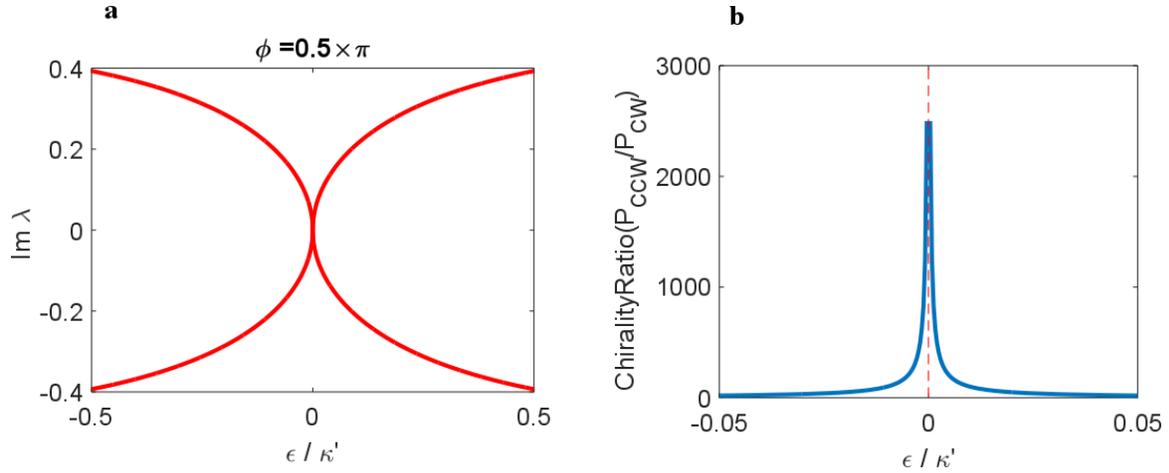

**Fig. S11| Robustness of the chiral mode lasing against defects. a**, Imaginary part of the eigenvalues of the counter-propagating modes in a microring with S-bend, in the presence of defects with strength $\epsilon$ in the coupling, when the optical phase of the S-bend is $\phi = \pi/2$. **b**, Power ratio of the respective CCW and CW components of the eigenvectors $V_{1,2}$. It is clear that a chiral mode (CCW) predominantly lases in the microring.

To show the broadband unidirectional behavior of the proposed technique, we also consider variation of the phase $\phi$ around $\pi/2$. Figure S12 depicts similar results for the eigenvalue spectra when $\phi = 0.4 \times \pi$ and $\phi = 0.6 \times \pi$. As it is clear from these plots, for a large range of $\phi$ (and hence wide range of wavelengths – especially those designed to be close to the lasing modes of the cavity) the spectrum always exhibits a nonzero decay difference between the two modes, regardless of the magnitude of the perturbation. In addition, unlike conventional systems involving regular exceptional points, this structure operates in a PT-symmetry broken phase for a broad range of variables, thus enforcing unidirectional lasing. This largely facilitates the direct generation of OAM beams from such devices in a controllable, yet broadband manner. Even in the ultimate case when $\phi = n\pi$, the modes, though showing splitting only in the real frequency domain, remain highly unidirectional as long as the perturbation is small in comparison to the intentional coupling ($\kappa'$) induced through the S-bend.

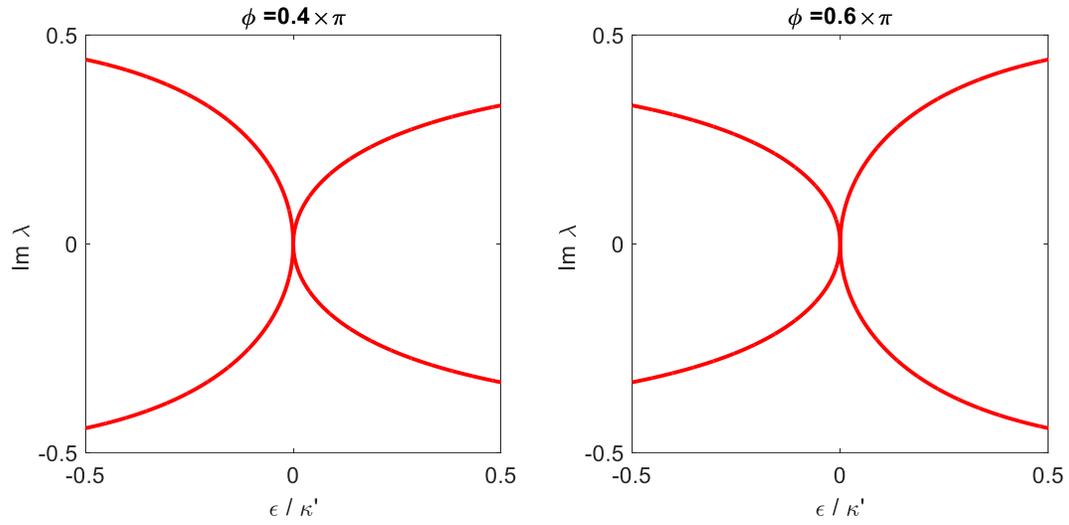

**Fig. S12| Broadband chiral behavior in microring with an S-bend.** Imaginary part of the eigenvalues $\lambda_{1,2}$ when the value of $\phi$ deviates around $\pi/2$.